\input vanilla.sty
\baselineskip=15 pt
\def\p{\partial}
\def\d{\delta}
\def\f{\frac}

\def\i{\infty}
\def\s{\sum }
\def\t{\tilde}

\def\e{\text {e}}

\def\C{\Cal }

\def\a{\alpha }
\def\res{\text {res}}
\def\th{\text {th}}
\def\({\left (}
\def\){\right )}
\def\[{\left [}
\def\]{\right ]}
\pagewidth {14cm}
\pageheight {19cm}
\TagsOnRight
\title Hamiltonian Structures of KdV-Type Hierarchies \\
and Associated $W$-Algebras \endtitle
\author Yi Cheng and Qing Chen  and Jingsong He\\  \   \\
{\it Department of Mathematics}\\
{\it University of Science and Technology of China}\\
{\it Hefei 230026, Anhui, People's Republic of China}\\
\endauthor
\heading ABSTRACT \endheading
\smallpagebreak
\vskip 0.4cm
{\narrower The $(n,m)^{\th }$ KdV hierarchy is a restriction of the KP hierarchy to a 
submanifold of pseudo-differential operators in a radio form. Explicit formula of the restricted Hamiltonian structure of KP is given which provides a new, more constructive proof of the isomorphism between the associated $W(n,m)$-algebra to $W_{n+m}\oplus W_m\oplus U(1)$ algebra, and the Hamiltonian property of the $(n,m)^{\th }$ KdV hierarchy as well as its Lax-Manakov triad representation. Similarly the Hamiltonian property for a version of modified $n^{\th }$ KdV and the isomorphism between $W_n$-algebra to $W_l\oplus W_m\oplus U(1)$ algebra are shown, where $l+m=n$. The role of $U(1)$ current in both cases is also explained.\par}
\newpage

Recently, Bonora et al. [1,2] proposed a KdV-type integrable system, which they called the 
$(n, m)^{\th }$ KdV hierarchy. This hierarchy is obtained by restricting the $n^{\th }$
order pseudo-differential operators (PDOs) associated with the KP hierarchy 
to a submanifold of PDOs of the following form
$$L=\e ^{-m\theta }A\e ^{-n\theta }B^{-1}\e ^{(n+m)\theta }, \tag 1 $$
where $m\geq 2$, $A$ and $B$ are two pure differential operators of the $(n+m)^{\th }$ 
and $m^{\th }$ order respectively and both without the second leading terms. Thus the 
hierarchy contains equations for the $n+2m-1$ independent unknowns: $\theta $ (or 
$J=\theta _x$), coefficients of $A$ and $B$. They showed the compatibility of the restriction with the KP hierarchy, displayed the first nontrivial flow and most importantly, they proved that the so-called $W(n,m)$-algebra appearing as the second Poisson bracket of the 
$(n, m)^{\th }$ KdV hierarchy is isomorphic via a Miura map to the direct sum of 
the $W_{n+m}$-algebra and $W_m$-algebra, as well as an additional $U(1)$ current algebra. As was shown in [2], the proof of this isomorphism depends on the derivation of the first nontrivial Hamiltonian equation and the uniqueness of the second Poisson structure of the equation. In [3] Dickey showed another viewpoint on this hierarchy and provided a more constructive proof of compatibility of such restriction with the KP hierarchy. \par
Despite of these studies and a closely connection of the $(n,m)^{\th }$ KdV hierarchy and
the so-called constrained KP hierarchy given in a series papers (see [4-8]), the structure of 
all equations in the $(n,m)^{\th }$ KdV hierarchy, particularly their Hamiltonian properties  
still need the further inverstigation. \par
In this letter we first show the restriction of the Poisson bracket of the KP hierarchy and
show the explicit formula of the Poisson bracket in terms of the finite numbers of coordiantes (or called fields) $\theta $, and coefficients of $A$ and $B$. As results we are able to present an exact and pure algebraic proof of the isomorphism between $W(n,m)$-algebra and $W_{n+m}\oplus W_m\oplus U(1)$ algebra, and show that all the equations in the $(n,m)^{\th }$ KdV hierarchy are Hamiltonian and possess the Lax-Manakov triad representations. \par
Then we immediately reconganize that the above results can be generalized to the $n^{\th }$ KdV hierarchy (or called Gelfand-Dickey hierarchy). If we restrict the $n^{\th }$ pure differential operator $L$ by
$$L=\e ^{-m\theta }A\e ^{n\theta }B\e ^{-l\theta }, \tag 2 $$
with $l+m=n$, $2\leq l, m\leq n-2$, and $A$ and $B$ being respectively of the $l^{\th }$ and $m^{\th }$ order of pure differential operators both without the second leading terms, then we have the parallel results, namely we can prove that the $W_n$-algebra associated with the second Hamiltonian structure of the $n^{\th }$ KdV hierarchy is isomorphic via a Miura map to the direct sum of a $W_l$-algebra, a $W_m$-algebra and a $U(1)$ current algebra, and we have a Hamiltonian system and the Lax-Manakov triad representation for $J=\theta _x, A, B$. Euqations in this system can be considered as a version of the modified equations of the $n^{\th }$ KdV equations.  \par
In both cases of $W(n,m)$-algebra and $W_n$-algebra, the $U(1)$ current $J$ plays an important role. We explain its role from the viewpoint of free fields realization and prove that it is
a conserved dentity of the $(n,m)^{\th }$ KdV hierarchy and the modified $n^{\th }$ KdV hierarchy for $J, A, B$. \par
\vskip 0.3cm
In general let
$$\Cal L=\p ^n+\s _{j=-\i }^{n-1}u_j\p ^j, \tag 3   $$
be the PDO of order $n$. The bi-Hamiltonian structures can be built 
on the space of such type PDOs, in particular the second Poisson bracket is given by (see [9])
$$\{\t f, \t g\}^{(n)}=\int \res \(\(\C L\f {\d f}{\d \C L}\)_+\C L-\C L
\(\f {\d f}{\d \C L}\C L\)_+\)
\f {\d g}{\d \C L}dx, \tag 4 $$
where $\t f=\int f(u_{n-1},u_{n-2},\cdots )dx$ with $f(u_{n-1},u_{n-2},\cdots )$ being
differential polynomial in $u_{n-1},u_{n-2},\cdots , $
$$\f {\d f}{\d \C L}=\sum _{r=-\i }^{n-1}\p ^{-r-1}\f {\d f}{\d u_r}, \tag 5  $$
and similarly for $\t g$ and $\d g/\d \C L$. Here and after for a PDO 
$A=\s _{i\leq n}a_i\p ^i$, $\res A=a_{-1}$ and $A_{\pm }$ denote its differential 
and residual parts respectively. \par
If we take the reduction of the Poisson bracket (4) to the submanifold of $u_{n-1}=0$, 
the term $\d f/\d u_{n-1}$ in $\d f/\d \Cal L$ becomes indefinite, so the 
following condition
$$\res \[\f {\d f}{\d \C L}, \C L\]=0 \tag 6 $$
should be teken into account such that $\d f/\d u_{n-1}$ is eliminated from (2) by
expressing it in terms of other coefficients $\d f/\d u_r$ [9]. 
By this condition we have \par
{\bf Proposition 1}. {\it The Poisson bracket reduced to the submanifold $u_{n-1}=0$ is 
in the form
$$\aligned \{\t f, \t g\}_R^{(n)}=&\int \res \(\(L\f {\d f}{\d L}\)_+L-L
\(\f {\d f}{\d L}L\)_+\)\f {\d g}{\d L}dx \\
&-\f 1n\int \(\res \[L, \f {\d g}{\d L}\]\(\p ^{-1}\res \[L, \f {\d f}{\d L}\]\)\)dx,
\endaligned \tag 7 $$
where 
$$L=\p ^n+\s _{j=-\i }^{n-2}u_j\p ^j, \tag 8 $$
and} 
$$\f {\d f}{\d L}=\sum _{r=-\i }^{n-2}\p ^{-r-1}\f {\d f}{\d u_r}. \tag 9 $$
Equation (7) was appeared in [1] as well as in [7] for a special case, however 
it still needs a more general proof. We display such proof below. \par
{\it Proof}. Let
$$X=\s _{r=-\i }^{n-1}X_r=\p ^{-n}X_{n-1}+\t X,$$
the conditions $u_{n-1}=0$ and (6) lead to
$$\res [\C L, X]=\res [L, \p ^{-n}X_{n-1}]+\res [L, \t X]=0,$$
from which we have
$$n\f {\p X_{n-1}}{\p x}+\res [L, \t X]=0. \tag 10 $$
Substituting $X=\d f/\d \C L=\p ^{-n}X_{n-1}+\t X$, $\t X=\d f/\d L$ and the similar
expression for $Y=\d g/\d \C L$ into (4) and noting that 
$$(XL)_+= X_{n-1} + (\t XL)_+, \ \ \ (LX)_+= X_{n-1} + (L\t X)_+,$$
we may directly have the result. \par
In terms of the fields $\{u_j\}$, the Poisson brackets among themself are given by
$$\{u_r(x), \ u_s(y)\}_R^{(n)}=J_{r,s}^{(n)}(u)\d (x-y), \tag 11 $$
for $r,s=n-2, \cdots $ and is called the $\hat W^{(n)}_{\i }$-algebra (see [10]), it contains
the following Virasoro algebra $Vir[c(n)]$
$$\{u_{n-2}(x), u_{n-2}(y)\}^{(n)}_R=-\(\f c{12}\p ^3+u_{n-2}\p+\p u_{n-2}\)\d (x-y), \tag 12 $$
as its subalgebra, where 
$$c(n)=(n-1)n(n+1) \tag 13 $$
is the central charge. \par
It is well-known (see [9]) that the flows in the KP hierarchy are Hamiltonian
with respect to the Poisson bracket (7) and the Hamiltonians are 
$$\t h_r(L)=\f nr\int \res L^{\f rn}dx, \ \ r=1,2,\cdots  \tag 14 $$
These quantities commute with each other. The $r^{\th }$ flow is written as
$$\aligned \p _{t_r}L&=\s _{j=-\i }^{n-2}\(\p _{t_r}u_j\)\p ^j \\
&=\s _{j=-\i }^{n-2}\{\t h_r, u_j\}\p ^j \\
&=\(L\f {\d h_r}{\d L}\)_+L-L\(\f {\d h_r}{\d L}L\)_+.
\endaligned $$
Since 
$$\f {\d h_r}{\d L}=L^{\f {r-n}n} \ \ \text {mod} \ R(-\i ,-n-1), \tag 15 $$
where $R(i,j)$ denotes the set of all PDOs of the form $P=\s _{r=i}^jP_r\p ^r$,
the $r^{\th }$ flow can be written as the following Lax representation
$$\p _{t_r}L=P_rL-LP_r, \tag 16 $$
with
$$P_r=\(L^{\f rn}\)_+. \tag 17 $$
\par
When $L$ is in the subspace consisting of pure differential operators of order $n$, (16) represents the $n^{\th }$ KdV hierarchy and (7) its second Poisson bracket. The correspondent Poisson algebra (11) is called $W_n$-algebra, where the subscript runs from $n-2$ to $0$.\par
\vskip 0.3cm
Now we restrict the PDO in the following form
$$L=\e ^{-m\theta }A\e ^{\a n\theta }B^{\a }\e ^{-\a l\theta },  \tag 18 $$
where 
$$\a =\pm 1, \ \ l=n - \a m, \tag 19 $$
with $2\leq m\leq n-2$, and $A$ and $B$ are two pure differential operators with the order of $l$ and $m$ respectively
$$\aligned &A=\p ^l+v_{l-2}\p ^{l-2}+\cdots +v_0, \\
           &B=\p ^m+w_{m-2}\p ^{m-2}+\cdots +w_0. 
\endaligned  \tag 20 $$
For $\a =-1$ or $\a =1$, (18) is the restriction (1) or (2). \par
So we have a group of new fields $\theta $ (sometime we use 
$J=\theta _x$), $v_{l-2}, \cdots , v_0$ and $w_{m-2}, \cdots , w_0$. The Miura map between the fields $u_{n-2}, \cdots , $ and the new fields can be obtained by comparing  coefficients in (18). According to these new fields we have \par
{\bf Porposition 2}. {\it The restriction of (18) leads the Poisson bracket in (7) to}
$$\aligned
\{\t f, \t g\}_R^{(n)}=&-\f {\a }{lmn}\int \f {\d f}{\d \theta }\(\p ^{-1}\f {\d g}{\d \theta }\)dx \\
&+\int \res \(\(A\f {\d f}{\d A}\)_+A-A
\(\f {\d f}{\d A}A)_+\)\)\f {\d g}{\d A}dx, \\
&+\f 1l\int \(\res \[A, \f {\d f}{\d A}\]\(\p ^{-1}\res \[A, \f {\d g}{\d A}\]\)\)dx \\
&+\a \int \res \(\(B\f {\d f}{\d B}\)_+B-B
\(\f {\d f}{\d B}B\)_+\)\f {\d g}{\d B}dx, \\
&+\f {\a }m\int \(\res \[B, \f {\d f}{\d B}\]\(\p ^{-1}\res \[B, \f {\d g}{\d B}\]\)\)dx.
\endaligned \tag 21 $$
\par
{\it Proof}. Here we only show the proof for $\a =-1$, in this case $l=n+m$. The other 
case ($\a =1$) is easier. First we calculate $\d \t f$ with respect to $L$ and $\theta , 
A, B$ respectively. We have
$$\aligned
\d \t f=&\int \res \f {\d f}{\d L}\d Ldx \\
=&\int \res \f {\d f}{\d L}\(-m\d \theta L+\e ^{-m\theta }\d A\e ^{-n\theta }
B^{-1}\e ^{l\theta } \right .\\
&-n\e ^{-m\theta }A\e ^{-n\theta }\d \theta B^{-1}\e ^{l\theta } \\
&-\e ^{-m\theta }A\e ^{-n\theta }B^{-1}\d BB^{-1}\e ^{l\theta } \\
&\left .+l\e ^{-m\theta }A\e ^{-n\theta }B^{-1}\e ^{l\theta }\d \theta \),
\endaligned $$
on the one hand and
$$\d \t f=\int \res \(\f {\d f}{\d A}\d Adx+\f {\d f}{\d B}\d B+
\f {\d f}{\d \theta }\d \theta \)dx $$
on the other hand. So compare coefficients of $\d A, \ \d B$ and $\d \theta $ in the 
above two expressions, we find
$$\aligned
\f {\d f}{\d \theta }=&\res \(-mL\f {\d f}{\d L}+l\f {\d f}{\d L}L 
-nB^{-1}\e ^{l\theta }\f {\d f}{\d L}\e ^{-m\theta }A\e ^{-n\theta }\),  \\
\f {\d f}{\d A}=&\e ^{-n\theta }B^{-1}\e ^{l\theta }\f {\d f}{\d L}\e ^{-m\theta }
 \ \ \ \ \ \ \text {mod} \ R(-\i ,-l)\cup R(0,\i ), \\
\f {\d f}{\d B}=&-B^{-1}\e ^{l\theta }\f {\d f}{\d L}\e ^{-m\theta }
A\e ^{-n\theta }B^{-1} \ \ \text {mod} \ R(-\i ,-m)\cup R(0,\i ).
\endaligned \tag 22 $$
Then the substitution of them and similar expressions associated with $\t g $ into the 
right hand side of (21) completes our proof. \par
Notice that in (21), the second term and the third term form the second Poisson
bracket associated with the $l^{\th }$ KdV hierarchy and the last two terms form the 
second Poisson bracket for the $m^{\th }$ KdV hierarchy, therefore the restriction 
decomposes the Poisson bracket. In terms of the fields $J$, $v_j$'s and $w_j$'s, 
this decomposition reads
$$\aligned 
&\{J(x), J(y)\}_R^{(n)}=\f {\a }{lmn}\d '(x-y), \\
&\{v_i(x), \ v_j(y)\}_R^{(n)}=J_{i,j}^{(l)}(v)\d (x-y), \\
&\{w_r(x), \ w_s(y)\}_R^{(n)}=\a J_{r,s}^{(m)}(w)\d (x-y),
\endaligned \tag 23 $$
and these three groups of fields mutually commute, where $1\leq i,j\leq l-2$
and $1\leq r,s\leq m-2$. So we have \par
{\bf Proposition 3}. {\it The $W(n,m)$-algebra and the $W_n$-algebra are isomorphic, respectively via a Miura map, to $W_l\oplus W_m\oplus U(1)$ algebra, where $l=n+m$, $m\geq 2$ for the isomorphism of $W(n,m)$-algebra, and $l=n-m$, $2\leq m\leq n-2$ for the isomorphism of $W_n$-algebra.}\par

The isomorphism for $W(n,m)$-algebra was shown in [2] and the proof there depends on the derivation of a nontrivial Hamiltonian equation and the uniqueness of the second Hamiltonian structure of this equation. Here we have provided a more constructive and pure algebraical proof not only for $W(n,m)$-algebra but also for $W_n$-algebra. \par
It is very interest, although seems obviously, that the subalgebra $Vir[c(n)]$ in (12) also has a similar decomposition, i.e. $Vir[c(n)]$ is decomposed as a direct sum of the subalgebra
$Vir[c(l)]$ of $W_l$ and the sublagebra $Vir[c(m)]$ of $W_m$, as well as a Virasoro algebra
$Vir[c_3]$ with central charge $c_3=\a 3lmn $. We explain this decomposition for $\a =1$, the other case is similar. By comparing coefficients on both sides of (2) (i.e. (18) with $\a =1$), we have
$$u_{n-2}=v_{l-2}+w_{m-2}+\sigma , \tag 24 $$
where
$$\sigma =-\f {lmn}2\(J_x+J^2\). \tag 25 $$
The fact can be shown immediately since $v_{l-2}$ and $w_{m-2}$ generate respectively the Virasoro algebras $Vir[c(l)]$ and $Vir[c(m)]$ and $\sigma $ also generates a Virasoro $Vir[c_3]$ with the central charge $c_3=3lmn$. \par
As was pointed in [2], the fields $J$ behaves like a gluon, which mediates the interaction between the $W_m$-algebra and $W_{n+m}$-algebra in the case of the $W(n,m)$-algebra. Here let us have a look of its behaviour from the viewpoint of the free field realization of the $W_n$-
algebra. Similar discussion is also valid in the case of $W(n,m)$-algebra. It is well-known that (see [11]) the factorization 
$$L(\p )=\p ^n+\sum _{j=0}^{n-2}u_j\p ^j=(\p +p_1)(\p +p_2)\cdots (\p +p_n),
 \tag 26 $$
with 
$$\s _{j=1}^np_j=0,  \tag 27 $$
leads to the Poisson brackets among the fields $p_j$ 
$$\{p_r(x), p_s(y)\}^{(n)}_R=(\d _{rs}-\f 1n)\d '(x-y), \ \ \ \ r,s=1,2,\cdots ,n.
\tag 28 $$
The  free field realization of $W_n$-algebra can be constructed by introducing $n-1$ free currents $j_1=\varphi '_1, \cdots , j_{n-1}=\varphi '_{n-1}$ and an overcomplete set
of vectors $\vec h_r, \ \ r=1,\cdots ,n$ in $(n-1)$-dimensional Euclidean space
with
$$\sum _{r=1}^n\vec h_r=0, \ \ \vec h_r \cdot \vec h_s=(\d _{rs}-\f 1n).
\tag 29 $$
Let $n=l+m$, $2\leq l,m\leq n-2$, and introduce a group of new fields by
$$\aligned
&J=\f 1{lm}\sum _{r=1}^lp_r=-\f 1{lm}\sum _{r=l+1}^{n}p_r, \\
&\bar p_r=p_r-mJ, \ \ r=1, \cdots ,l, \\
&\bar p_r=p_r+lJ, \ \ r=l+1, \cdots ,n=l+m.
           \endaligned \tag 30 $$
Then we find that
$$\sum _{r=1}^l\bar p_r=\sum _{r=l+1}^{n}\bar p_r=0. \tag 31 $$
The Poisson brackets among these new fields are
$$\aligned &\{J(x), J(y)\}^{(n)}_R=\f 1{lmn}\d '(x-y), \\
           &\{\bar p_r(x), \bar p_s(y)\}^{(n)}_R=(\d _{rs}-\f 1l)\d '(x-y),
           \ \ r,s=1, \cdots ,l, \\
           &\{\bar p_k(x), \bar p_l(y)\}^{(n)}_R=(\d _{kl}-\f 1m)\d '(x-y),
           \ \ k,l=l+1, \cdots , n=l+m, \endaligned \tag 32 $$
and all other possible brackets vanish. So the field $J$ decomposes the algebra (28) into two
pieces of the same type of algebra. \par
Let 
$$\aligned &A(\p )=(\p +\bar p_1)\cdots (\p +\bar p_l), \\
           &B(\p )=(\p +\bar p_{l+1})\cdots (\p +\bar p_n), \endaligned  \tag 33 $$
they are in the form of (20). By using (30) we have the relation between $L$, and $A$ and $B$
$$L=A(\p +mJ)B(\p -lJ), \tag 34 $$
which is the same as (2). \par
To study the Hamiltonian property of the equations for $J=\theta _x$, $A$ and $B$, we first define
$$\aligned 
&\hat L=\e ^{\a l\theta }B^{\a }\e ^{-\a n\theta }A\e ^{m\theta }, \\
&\hat P_r=\(\hat L^{\f rn}\)_+,
\endaligned \tag 35 $$
then we have \par
{\bf Proposition 4}. {\it The restriction of $\t h_r(L)$ on the subspace of PDOs in the form of
(18) is the Hamiltonian of the flow for $J=\theta _x, A, B$. The flow can be represented by 
the Lax-Manakov triad representation.
$$\aligned
&\p _{t_r}J=\f1{ml}\(\res \hat L^{\f rn}-\res L^{\f rn}\), \\
&\p _{t_r}A=M_rA-A\hat M_r, \\
&\p _{t_r}B=N_rB-B\hat N_r,
\endaligned \tag 36 $$
where
$$\aligned
&M_r=m\theta _{t_r}+\e ^{m\theta }P_r\e ^{-m\theta }, \\
&\hat M_r=m\theta _{t_r}+\e ^{m\theta }\hat P_r\e ^{-m\theta }, 
\endaligned \tag 37 $$
and in the case of $\a =-1$, $l=n+m$:
$$\aligned 
&N_r=l\theta _{t_r}+\e ^{l\theta }P_r\e ^{-l\theta }, \\
&\hat N_r=l\theta _{t_r}+\e ^{l\theta }\hat P_r\e ^{-l\theta },
\endaligned \tag 38 $$
while in the case of $\a=1$, $l=n-m$:}
$$\aligned 
&N_r=-l\theta _{t_r}+\e ^{-l\theta }\hat P_r\e ^{l\theta }, \\
&\hat N_r=-l\theta _{t_r}+\e ^{-l\theta } P_r\e ^{l\theta }.
\endaligned \tag 39 $$

{\it Proof}. We also show the proof for $\a =-1$ and $l=n+m$. Because of (15) we have
$$\f {\d h_r}{\d \theta }=\res \(nL^{\f rn}-nB^{-1}\e ^{l\theta }
L^{\f {r-n}n}\e ^{-m\theta }A\e ^{-n\theta }\),$$
the second term in the above can be written as 
$$\aligned 
\(B^{-1}\e ^{l\theta }L^{\f {r-n}n}\e ^{-m\theta }A\e ^{-n\theta }\)^{n\f 1n}=
&\(B^{-1}\e ^{l\theta }L^{r-1}\e ^{-m\theta }A\e ^{-n\theta }\)^{\f 1n} \\
&=\(\e ^{l\theta }\hat L^r\e ^{-l\theta }\)^{\f 1n} \\
&=\e ^{l\theta }\hat L^{\f rn}\e ^{-l\theta },
\endaligned $$
so
$$\f {\d h_r}{\d \theta }=n\(\res L^{\f rn}-\res \hat L^{\f rn}\).$$
The equation for $\theta $ is 
$$\aligned 
\p _{t_r}\theta &=\{\t h_r, \theta \} \\
&=-\f 1{lmn}\(\p ^{-1}\f {\d h_r}{\d \theta }\), \endaligned $$
which immediately implies the first equation in (36). \par
To calculate the equation for $A$, we first have 
$$A\f {\d h_r}{\d A}=\e ^{m\theta }L^{\f rn}\e ^{-m\theta },$$
and 
$$\aligned 
\f {\d h_r}{\d A}A&=\e ^{-n\theta }B^{-1}\e ^{l\theta }L^{\f {r-n}n}\e ^{-m\theta }A \\
&=\(\e ^{-n\theta }B^{-1}\e ^{l\theta }L^{\f {r-n}n}\e ^{-m\theta }A\)^{n\f 1n} \\
&=\(\e ^{-n\theta }B^{-1}\e ^{l\theta }L^{r-1}\e ^{-m\theta }A\)^{\f 1n} \\
&=\(\e ^{m\theta }\hat L^r\e ^{-m\theta }\)^{\f 1n}=\e ^{m\theta }\hat L^{\f rn}
\e ^{-m\theta },
\endaligned $$
so 
$$\res \[A, \f {\d h_r}{\d A}\]=\res L^{\f rn}-\res \hat L^{\f rn}=-lm\theta _{xt_r}.$$
The equation for $A$ now reads
$$\aligned \p _{t_r}A=&\{h_r, A\}=\s _{j=0}^{l-2}\{h_r, v_j\}\p ^j \\
=&\(A\f {\d h_r}{\d A}\)_+A-A\(\f {\d h_r}{\d A}A\)_+ -ml\theta_{t_r,x}\partial^{l-1}\\
&+m\s _{j=0}^{l-2}\(\int \(\theta _{t_r}\res \[A, \f {\d v_j}{\d A}\]\)dx'\)\p ^j,
 \endaligned $$
where $\d v_j/\d A=\p ^{-j-1}\d (x'-x)$. The last two terms in the above is
$m\(\theta _{t_r}A-A\theta _{t_r}\),$
so we have the second equation in (36) and similarly the third one.
\par
Let us show the first nontrivial flow ($r=2$) with $\a =\pm 1$ and $l=n-\a m$ 
$$\aligned 
\p _{t_2}J=&\a \(\f 2{nl}v-\f 2{nm}w+(l-\a m)J^2\)_x,\\
\p _{t_2}A=&\(\(\p -mJ\)^2+\f 2lv-mlJ_x-m^2J^2\)A\\
&-A\(\(\p -mJ\)^2+\f 2lv+mlJ_x-m^2J^2\),\\
\p _{t_2}B=&\(\(\p +\a lJ\)^2+ \f 2mw+\alpha mlJ_x-l^2J^2\)B\\
&-B\(\(\p +\a lJ\)^2+\f 2mw-\alpha mlJ_x-l^2J^2\),
\endaligned \tag 40 $$
where we have rename $v_{l-2}$ by $v$ and $w_{m-2}$ by $w$. In the case of $\a =-1$ and $l=n+m$, the above equation is exactly the same as the equation shown in [2].
\par
{\bf Proposition 5}. {\it Define
$$\t J=\int Jdx, \tag 41 $$
then it is the conserved quantity of the hierarchy in (36), and commutes
with other conserved quantities $\t h_r$.} \par
{\it Proof:} First using the relation
$$\hat L=\e ^{-m\theta }A^{-1}\e ^{m\theta }L\e ^{-m\theta }A\e ^{m\theta },$$
for $L$ in (18) and $\hat L$ in (35), then the identity $\int \res PQdx=
\int \res QPdx$ for any PDOs $P$ and $Q$ immediately implies that 
$$\f {\p \t J}{\p t_r}=\f {\p }{\p t_r}\int Jdx=0.$$
On the other hand one can check that  
$$\{\t h_r, \t J\}_R^{(n)}=0,$$
for the Poisson bracket in (21). 
\par
{\it Remark}. It can easily checked that both $L$ in (18) and $\hat L$ in (35) satisfy 
the KP hierarchy when $\a =-1$ and $n^{\th }$ KdV hierarchy when $\a =1$. In particular when $\a =1$, equations in (36) are type of modified ones of the $n^{\th }$ KdV hierarchy.
\vskip 0.4cm
\subheading {Acknowledgements } \par
Part of this work has been done during the frist author's visit in the City University of Hong Kong, he would like to thank the Research Grants Council of Hong Kong for financial support.
The work was supported by the National Science Fund and the National
Basic Research Project for ``Nonlinear Science'' . \par
\newpage
\vskip 0.4cm
\subheading {References} \par
\item {[1]} L. Bonora and C. S. Xiong,  J. Math. Phys. {\bf 35}, 5781 (1994).
\item {[2]} L. Bonora, Q. P. Liu and C. S. Xiong, Commun. Math. Phys.
{\bf 175}, 177 (1996).
\item {[3]} L. A. Dickey, Lett. Math. Phys. {\bf 35}, 229 (1995).
\item {[4]} B. Konopelchenko, J. Sidorenko and W. Strampp, Phys. Lett. {\bf A157}, 17 (1991).
\item {[5]} Y. Cheng and Y. S. Li, Phys. Lett. {\bf A157}, 22 (1991).
\item {[6]} Y. Cheng, J. Math. Phys. {\bf 33}, 3774 (1992).
\item {[7]} Y. S. Wu and F. Yu, Phys. Rev. Lett. {\bf 66}, 2996 (1992).
\item {[8]} H. Aratyn, L. A. Ferreira, J. F. Gommez and A. H. Zimerman, J. Math. Phys. 
{\bf 36}, 3419 (1995).
\item {[9]} L. A. Dickey, {\it Soliton Equations and Hamiltonian Systems}, Advanced Series in Math. Phys. {\bf vol.12} (World Scientific 1991). \par
\item {[10]} J. M. Figueroa-O'Farrill, J. Mas and E. Ramos, Commun. Math. Phys. {\bf 158}, 17 (1993).
\item {[11]} I. Bakas, in {\it Nonlinear Fields: Classical, Ramdom, Semiclassical}, P. Garbaczewski and Z. Popowitz eds. (World Scientific, Singapore 1991), 2
\bye